\documentclass[12pt, preprint]{aastex}
\usepackage[breaklinks,colorlinks, 
urlcolor=blue,citecolor=blue,linkcolor=blue]{hyperref}
\usepackage{graphicx}	% For figures
\usepackage{natbib}	% For citep and citep
\usepackage{amsmath}	% for \iint
\usepackage{mathtools}
\usepackage{bm}
\usepackage[breaklinks]{hyperref}	% for blackboard bold numbers
\usepackage{hyperref}
\usepackage{algorithmic,algorithm}
\hypersetup{colorlinks}
\usepackage{color}
\usepackage{morefloats}
\definecolor{darkred}{rgb}{0.5,0,0}
\definecolor{darkgreen}{rgb}{0,0.5,0}
\definecolor{darkblue}{rgb}{0,0,0.5}
\hypersetup{ colorlinks,
linkcolor=darkblue,
filecolor=darkgreen,
urlcolor=darkred,
citecolor=darkblue}

\DeclareMathOperator*{\argmax}{arg\,max}
\newcommand{\beq}{\begin{equation}}
\newcommand{\eeq}{\end{equation}}

\newcommand{\etal}{et~al.}

\begin{document}

\author{
  Mohammadjavad~Vakili\altaffilmark{1},
  David~W.~Hogg\altaffilmark{1,2,3}}
\altaffiltext{1}{Center for Cosmology and Particle Physics, Department of Phyics,
             New York University, 4 Washington Pl., room 424, New York, NY, 10003, USA}
\altaffiltext{2}{Max-Planck-Institut f\"ur Astronomie, K\"onigstuhl 17, D-69117 Heidelberg, Germany}
\altaffiltext{3}{Center for Data Science, New York University, 726 Broadway, 7th Floor, New York, NY 10003, USA}
\email{mjvakili@nyu.edu}

\title{Do fast stellar centroiding methods saturate the Cram\'{e}r-Rao lower bound?}

\begin{abstract}
One of the most demanding tasks in astronomical image processing---in terms of precision---is 
the centroiding of stars. Upcoming large surveys are going to take images of 
billions of point sources, including many faint stars, with short exposure times. 
Real-time estimation of the centroids of stars is crucial for real-time PSF estimation, 
and maximal precision is required for measurements of proper motion. 

The fundamental Cram\'{e}r-Rao lower bound sets a limit on the root-mean-squared-error 
achievable by optimal estimators. In this work, we aim to compare 
the performance of various centroiding methods, in terms of saturating the bound, when they 
are applied to relatively low signal-to-noise ratio unsaturated stars assuming zero-mean 
constant Gaussian noise. In order to make this comparison, we present the ratio of the root-mean-squared-errors of 
these estimators to their corresponding Cram\'{e}r-Rao bound as a function of the signal-to-noise ratio 
and the full-width at half-maximum of faint stars. 

We discuss two general circumstances in centroiding of faint stars: (i) when we have a good estimate
of the PSF, (ii) when we do not know the PSF. In the case that we know the PSF, 
we show that a fast polynomial centroiding after smoothing the image by the PSF can be 
as efficient as the maximum-likelihood estimator at saturating the bound. 
In the case that we do not know the PSF, we demonstrate that although polynomial centroiding is not as optimal as PSF profile fitting, it comes very close to saturating the Cram\'{e}r-Rao lower bound in a wide range of conditions. We also show that the moment-based method of center-of-light 
never comes close to saturating the bound, and thus it does not deliver reliable estimates of centroids.    
\end{abstract}

\keywords{methods: statistical --- methods: data analysis --- techniques: image processing}

\section{Introduction}

Accuarate estimates of the centers of point sources, which are convolved with telescope point spread function (PSF), and atmospheric PSF in case of ground based telescopes, and the pixel response function, are crucial to further steps of
astronomical image processing. For instance, proper measurement of the shapes of galaxies
requires interpolation of the PSF estimates from the positions of stars across the
image to the positions of galaxies. At the position of each star, the PSF is estimated by sub-pixel 
shifting of the star so that the PSF is centered on its centroid. If the sub-pixel shifts are wrong, then 
the PSF estimates will be biased. Moreover, measurements of the parallaxes and the proper motions of stars
depend on how well we can measure their centroids. 

Ideally, we want a centroiding procedure that provides measurements as precise 
as possible without putting a huge computational burden on the photometric pipeline.
Reducing the computational cost becomes even more important in large surveys,
where we want to estimate the centroids of thousands of point sources detected
on the telescope's focal plane, for various real-time applications. 

The Cram\'{e}r-Rao lower bound (CRLB) sets a lower limit on the root-mean-squared error of estimators. When the root-mean-squared error arsing from an estimator approaches the bound, the bound is saturated by that estimator. In this paper, we study the optimality of various techniques for centroiding faint, unsaturated stars. Our requirement for optimality is saturation of the 
theoretically-set lower bound, known as the Cram\'{e}r-Rao lower bound, by the 
centroiding methods considered in this study. 

We apply a number of 
centroiding methods to a large number of simulated faint stars, assuming uncorrelated 
Gaussian noise, with different signal-to-noise ratio and size realizations. 
The Cram\'{e}r-Rao lower bound has an inverse relation with the signal-to-noise-ratio 
of stars. In the context of astrometry, the Cram\'{e}r-Rao lower bound saturation for least-squares estimators has been tested in specific limits in which the centroiding bias is negligible (\citealt{lobos}). 

Saturating the Cram\'{e}r-Rao lower bound in estimating the centroids of stars however, is limited
by the lack of knowledge about the exact shape of the PSF and presence of noise. 
There are many sources of noise such as the CCD readout noise, sky noise, errors resulting from 
incorrect flatfield corrections, and photon noise from the astronomical object itself. 
In this study, we limit our investigation to the simulated images that contain non-overlapping faint sources that are sky-limited. 

We focus the scope of this investigation to sky-limited images for which the sky level has been subtracted. Furthermore, we assume that any instrument gain has been calibrated out and that the 
simulated images are free of any contamination by cosmic rays, stray light from neighboring fields, or 
any other type of defect in real images. We expect these defects to move the centroiding errors further from the 
fundamental bound. We intend to investigate whether fast centroiding estimates can saturate the bound 
in a realistic range of low signal-to-noise ratio images that are sky-limited.

%To date, a number of softwares packages have been designed for the purpose of extracting astronomical
%sources and making catalogs. One of these softwares is SExtractor (\citealt{sextractor}),
%whose centroiding method involves first finding the zeroth moment of the object
%as a first-order estimate, and then iteratively correcting the centroid by computing
%the zeroth order moment of the object weighted by a Gaussian window function,
%until the correction falls below a particular threshold value.
%The width of the Gaussian window function is set by the object's half-light radius.
%Other examples are DAOPHOT (\citealt{daophot}), and DOPHOT (\citealt{dophot})
%which both assume analytic models for the stellar PSF profiles with centroid
%coordinates being free parameters of these models.
%DAOPHOT (DOPHOT) finds the centroids by fitting a Gaussian (a truncated power series) to 
%the light profile of stars.

Given an analytic expression for the PSF model adopted in this study,
we derive an expression for the fundamental lower bound on the centroiding error as
a function of the parameters of the PSF model (e.g. PSF size),
and signal-to-noise-ratio of stars. We create two sets of simulations for which we can 
compute the CRLB, one with variable signal-to-noise ratio and constant full width at half maximum (FWHM), and one 
with variable FWHM and constant signal-to-noise ratio. After applying
different centroiding methods to the simulations, we investigate how close 
these methods can get to saturating the CRLB for various ranges of background 
Gaussian noise level and PSF FWHM.

In this work, we focus on four centroiding methods. The first method is the maximum-likelihood 
estimator which involves fitting a PSF profile, assuming that we have a good PSF estimate, to the star. 
The second method estimates the centroid of a star by fitting a 2d second-order polynomial to 
the 3$\times$3 patch around the brightest pixel of the image after convolution with the PSF. 
The third method centroids stars by
 smoothing the image of stars by a Gaussian kernel of a fixed size,
 and then applying the same 3$\times$3 polynomial trick to the smooth
 image. This method is fast and does not require any knowledge of the 
PSF. The last method we consider, is a center-of-light centroiding 
(measurement of a first moment), applied to the 7$\times$7 patch around the brightest pixel of the image.

This paper is structured as follows. In Section \ref{sec:CRLB},
we discuss the Cram\'{e}r-Rao lower bound and derive
an analytic expression for the lower bound on the centroiding error
of the simulated data. 
In Section \ref{sec:method} we give a brief overview of 
centroiding methods used in our investigation.
In Section \ref{sec:data} we discuss the Cram\'{e}r-Rao lower bound satuaration
tests and their corresponding simulated data.
In Section \ref{sec:result}, we compare the performances of the methods
discussed in Section \ref{sec:method} with the CRLB derived in Section \ref{sec:CRLB}. Finally, we discuss and conclude in Section \ref{sec:discussion}.               

%%%%%%%%%%%%%%%%%%%%Cram\'{e}r-Rao lower Bound %%%%%%%%%%%%%%

\section{The Cram\'{e}r-Rao lower bound}\label{sec:CRLB}

The Cram\'{e}r-Rao lower bound sets a limit, in some sense, on how well a measurement 
can be made in noisy data.  The bound can only be computed in the context of a 
generative model, or a probabilistic forward model of the data. That is, we can 
only compute the CRLB in the context of assumptions about the properties of the data. 
However, it makes sense for us to use centroiding methods that saturate the CRLB under 
some reasonable assumptions, even if we find that those assumptions are not strictly correct in real situations.

The closer an estimator is to saturating the CRLB, the more information about the quantity that we 
need to estimate is preserved. The closer the root-mean-squared-error (RMSE) of a given estimator is to the bound,  
the more optimal---in terms of preseving the information---the estimator is. 

The Cram\'{e}r-Rao inequality \citep{crlb} sets a lower bound on the 
root-mean-squared error of unbiased estimators. The CRLB is given by the square-root of the inverse of 
the Fisher information matrix $\mathcal{F}$. Thus, in order to find the CRLB, it is sufficient to compute the Fisher matrix. 
This computation relies on a set of assumptions:

\begin{itemize}
%  \item Known, constant model observable with known dependence on the model parameters. 
%        In this work the model observables are the presumed known Moffat PSF profiles, 
%        and the model parameters are the centroids. 
  \item Known PSF model. In this work the presumed model is the Moffat PSF profile.
  \item Known, stationary noise process. In the context of centroiding stars, this is equivalent to 
        having background limited noise from sky background and CCD readout noise.
  \item Images are calibrated correctly. Flat-field is correctly calibrated.       
  \item Uncorrelated Gaussian noise with no outliers. 
\end{itemize}

%\todo{HOGG HELP: can you read the following paragraph and see whether it makes sense or not?}

Note that in this study, we explicitly focus on sky-limited images. In the sky-limited images, the contribution 
to the Poisson pixel noise is largely dominated by the sky rather than the objects. In sky-limited images, when the number of photons per pixel is large, the Poisson noise can be approximated by a Gaussian distribution. Therefore, the Gaussian noise assumption is only an approximation to the Poisson noise. This is a good approximation for a large set of astronomical images. 

A number of factors can produce correlation between pixels. These include detector imperfection, saturation, and post-processing of images such as smoothing, rotating, and shifting the images. In raw unsaturated images, pixel noise is close to uncorrelated. Instrument gain can introduce heteroscedasticity. In that case, the noise variance varies between pixels. In an upcoming publication on the inference of the HST WFC3-IR channel PSF (Vakili $\etal$, in preparation), we discuss proper treatment of centroiding in the presence of gain. For simplicity, we assume that per-pixel uncertainty remains constant across all pixels. 

%After subtraction of the 
%sky background level, distribution of the Poisson pixel noise can be approximated by a Gaussian %distribution with 
%zero mean. \todo{Also mention in the discussion: The assumption of uncorrelated Gaussian noise is only %an approximation 
%to a Poisson distribution which has broader tails than what we expect from a Gaussian!} 
%\todo{HOGG HELP: can you write a few words about uncorrelated noise?}

Let us assume that there are $M$ observables $\mathbf{f} = (f_{1}, ... , f_{M})$, each
related to $B$ model parameters $\boldsymbol{\mathbf{\theta}} = (\theta_{1} , ... , \theta_{B})$ 
\beq
f_{m} = f_{m}(\theta_{1} , ... , \theta_{B}).
\label{genmodel}
\eeq

Assuming uncorrelated Gaussian error with variance $\sigma^{2}_{m}$ for each observable $f_{m}$, elements
of the $B\times B$ Fisher matrix $\mathcal{F}_{ij}$ are given by
\beq
\mathcal{F}_{ij} = \sum_{m=1}^{M}\frac{1}{\sigma_{m}^{2}}\frac{\partial f_{m}}{\partial \theta_{i}}\frac{\partial f_{m}}{\partial \theta_{j}}
\label{fisher}
\eeq
%Let us assume that, for each parameter $\theta_{i}$, there exists a set of 
%asymptotically unbiased estimators $\{\hat{\theta}_{i}\}$. 
Let us assume that we have computed the root-mean-squared error on the parameter $\theta_{i}$ arising from applying an estimator to a large number of data. The Cram\'{e}r-Rao inequality states that this root-mean-squared error is greater than or equal to the $i$-th diagonal element of the inverse of the Fisher information matrix:
\beq
%\text{RMSE}[\{\hat{\theta_{i}}\}] \geq \sqrt{[\mathcal{F}^{-1}]_{ii}},
\text{RMSE} \geq \sqrt{[\mathcal{F}^{-1}]_{ii}},
\label{inequality}
\eeq
where the left hand side of the inequality is called the Cram\'{e}r-Rao bound on 
the root-mean-squared error of estimating the parameter $\theta_{i}$. Note that 
the bound is computed assuming that the model (equation~\ref{genmodel}) generating the data 
is known, and that uncertainties are given by additive uncorrelated Gaussian noise.

Based on Cram\'{e}r-Rao inequality (\ref{inequality}), \citet{crlb} defines efficiency of optimal estimators as the ratio of the CRLB and the root-mean-squared-error such that the maximum efficiency 
achievable by an estimator is unity. The closer the RMSE to the CRLB, 
the more information about the parameter of interest is preserved, and thus the more efficient the estimator is. 

Let us consider the case of a maximum likelihood estimate $\boldsymbol{\mathbf{\theta}}_{\text{ML}}$, 
where the likelihood function corresponds to the same generative assumptions that we used to compute the CRLB.

\begin{eqnarray}
\boldsymbol{\mathbf{\theta}}_{\text{ML}} &=& \argmax \mathcal{L}, \\
-2\ln \mathcal{L} &=& \sum_{m}\frac{1}{\sigma_{m}^{2}}( y_{m} - f_{m}(\boldsymbol{\mathbf{\theta}}))^{2}, \\
\end{eqnarray}
where $y_{m}$ is the $m$th component of the observed data $\mathbf{y}$
\beq
\mathbf{y} = \mathbf{f}(\boldsymbol{\mathbf{\theta}}_{\text{true}}) + \mathbf{n}.
\eeq

Maximum likelihood estimators can achieve maximum efficiency. That is, when a maximum likelihood %estimators are applied to a large number of data, the RMSE is greater than or equal to the CRLB (see
estimator is applied to a large number of data and RMSE is computed, the RMSE approaches the CRLB (see
\citealt{crlb}; \citealt{lecam} for proof) in which case the CRLB is saturated. Therefore, we want to investigate the conditions under which the RMSE arising from a given fast centroiding method is close to the CRLB, or whether it can saturate the CRLB.

%However, the relation (\ref{inequality}) does not necessarily hold for biased estimators. That is, 
%the root-mean-squared-error for a biased estimator can be smaller than the CRLB (see \citealt{lecam} %for examples).

%, or whether it can drop below the CRLB in which case the method is \emph{beating} the bound (and %therefore the estimator must be biased).   
  
In this investigation, the model observables for the noisy data are the pixel-convolved PSF (PSF profile evaluated at different pixel locations), and  
the model parameters under consideration are the centroid coordinates. Therefore, $\mathcal{F}$
is a 2$\times$2 matrix whose elements are given by

\beq
  \mathcal{F}_{ij} = \sum_{m}\frac{1}{\sigma^{2}}
                \frac{\partial f_{m}}{\partial \theta_{i}}\frac{\partial f_{m}}{\partial \theta_{j}},
\label{fish}
\eeq
where the summation is over pixels, $f_{m}$ is the value of the PSF at pixel location $m$,
$\theta=\{x_{c},y_{c}\}$, and $\sigma^{2}$ is variance of the uncorrelated Gaussian noise map $n(\mathbf{x}_{m})$
\begin{eqnarray}
\mathbb{E}[n(\mathbf{x}_{m})] &=& 0, \\
\mathbb{E}[n(\mathbf{x}_{m})n(\mathbf{x}_{m^{\prime}})] &=& \sigma^{2}\delta_{m,m^{\prime}}. 
\end{eqnarray}

Derivation of an explicit expression for the Fisher matrix $\mathcal{F}$ requires 
specifying a presumed correct PSF model.
We use the Moffat profile \citep{moffat} for our PSF simulations. 
The Moffat profile is an analytic model for stellar PSFs. It has broader wings than
a simple Gaussian profile. The surface brightness of the Moffat profile is given by
\beq
I(r) = \frac{F(\beta -1)}{\pi \alpha^{2}}[1+(r/\alpha)^{2}]^{-\beta},
\label{mof}
\eeq
where $F$ is the total flux, $\beta$ is a dimensionless parameter, and $\alpha$ is
the scale radius of the Moffat profile, with FWHM (hereafter denoted by $\gamma$)
being $2\alpha\sqrt{2^{1/\beta}-1}$. The Moffat PSF profile has been used in the PSF modeling required for weak lensing galaxy shape measurements (see \citealt{im3shape,im3shape_code}). It has also been used as one of the methods for generation of the PSF in simulation of images needed for weak lensing systematic studies (\citealt{galsim}).
At a fixed $\gamma$, Moffat profiles with lower values
of $\beta$ have broader tails. It is also important to note that for sufficiently large values of the 
parameter $\beta$, the Moffat PSF becomes arbitrarily close to a simple Gaussian PSF. 

Note that in our data generation, we simulate images (in the pixel space) that are Nyquist-sampled or close to Nyquist-sampled. All pixels in the images are identical, and the stars are simulated 
by sampling from the pixel-convolved PSF. In well-sampled images, the center of 
the pixel-convolved PSF must be very close to the center of the optical PSF.
 
In order to investigate the performance of centroiding methods for
 different background noise levels and different
values of the parameter $\gamma$, simulation of a large number of images of stars---for which the exact positions of centroids
and their corresponding lower bounds are known---is required.

Given the PSF model (\ref{mof}), an expression for the CRLB as a function of the size, and SNR of stars can be 
derived. For further simplicity, the flux of all stars in our simulations are set to unity and per-pixel 
uncertainties are assumed to be uncorrelated Gaussian.

Moreover, it is more convenient to work with the signal-to-noise ratio
(SNR) instead of the variance of the Gaussian noise.
We use the definition of SNR according to which SNR is given by the ratio
 of the mean and variance of the distribution
which the flux estimator is drawn from. Assuming that the total flux from
the point source is $F$, and that the sub-pixel shifted PSF at the $i$-th pixel is given
by $P_{i}$. Therefore the brightness of the $i$-th pixel $y_{i}$ is drawn from
a Gaussian distribution 
\beq
p(y_{i}) = \mathcal{N}(FP_{i},\sigma^{2}). 
\eeq

The optimal estimator of flux is the matched-filter flux estimator 
$\tilde{F}=\sum_{i}y_{i}P_{i}$. It can be shown that 
\beq
p(\tilde{F}) = \mathcal{N}(F , \frac{\sigma^{2}}{\sum_{i}P_{i}^{2}}),
\eeq  
which leads us to
\beq
\begin{array}{l}
\text{SNR} = \frac{F\sqrt{\sum_{i} P_{i}^{2}}}{\sigma}.
\end{array}
\label{snr}
\eeq

 In the case of Moffat profiles (\ref{mof}) with total flux of stars set to unity, 
the SNR given in (\ref{snr}) can be analytically 
expressed in terms of the per pixel uncertainty
$\sigma$, FWHM $\gamma$, and also $\beta$, the dimensionless parameter of (\ref{mof})
\beq
\text{SNR} = \frac{2(\beta-1)(2^{1/\beta}-1)^{1/2}}{\pi^{1/2}(2\beta-1)^{1/2}}\frac{1}{\sigma \gamma}.
\label{snr2}
\eeq

Equation (\ref{snr2}) implies that at a fixed $\gamma$ and background Gaussian noise 
with variance $\sigma^{2}$, stars with broader tails (lower $\beta$) have a lower SNR.
On the other hand, stars with higher $\beta$ have higher SNR. 
For sufficiently large $\beta$---where the PSF can be
approximated by Gaussian profile---SNR is approximately given by $0.664/(\sigma\gamma)$.
Furthermore, at a fixed $\beta$ and variance of the background noise $\sigma^{2}$,
observed stars with higher $\gamma$ have lower SNR.  

Throughout this investigation, $\beta$ is held fixed at the fiducial value of $\beta=$ 2.5, where SNR
is given by the following expression
\beq
\text{SNR} \simeq \frac{0.478}{\sigma \gamma}\;\;\; \mathrm{for}\;\;\; \beta = 2.5.
\eeq

Given the analytic expression for the Moffat PSF model (\ref{mof}), and choice of $\beta=2.5$, 
the inverse of the Fisher matrix is given by
\beq
  \mathcal{F}^{-1} \simeq \Big(0.685 \frac{\gamma}{\text{SNR}}\Big)^{2} 
  \begin{pmatrix}
      1 & 0\\
      0 & 1\\
  \end{pmatrix}.
\label{crlbmoffat}
\eeq

Equation (\ref{crlbmoffat}) implies that at given SNR and $\gamma$,
CRLB for each component of centroid is approximately given by $0.685\gamma/\text{SNR}$,
and that a good centroiding technique delivers centroids with
root-mean-squared-error close to this. 

It is worth noting that for any PSF model whose radial light profile is some function of 
$r/\gamma$, CRLB has the same functional form, in that it is proportional to the ratio
between $\gamma$ and the SNR. 
For PSF profiles with shorter tails (e.g., Gaussian), the prefactor of 0.685 in (\ref{crlbmoffat})
becomes smaller. In the particular case of Gaussian PSF, the prefactor is approximately 0.6. 

\section{Centroiding methods}\label{sec:method}

In this section, we briefly discuss the approximate and the 
non-approximate centroiding methods considered in this 
study. The first two methods require knowledge of the PSF at the position of star. That is, 
the shape and the size of the PSF is known and the only unknown variables are the coordinates of 
the centroids of stars. Note that in practice however, size and shape of the PSF are also estimated along with the centroid. In the following, we assume that the size and shape of the PSF are known. For the last two methods, we do not use any information about the PSF. 

\subsection{Centroiding by fitting a correct PSF profile}

We examine fitting an exact PSF profile to the stars. That is, 
in our Cram\'{e}r-Rao bound saturation tests, we find the best
estimates of flux and centroid by maximizing the likelihood using 
the correct PSF model. In the model, the size of the Moffat PSF is 
assumed to be correct. We expect this method to perform best in 
determining the centroids of stars, and deliver RMSE equal to 
Cram\'{e}r-Rao bound.

\subsection{Matched-filter polynomial centroiding}

Let us consider the case in which we have a good estimate of the pixel-convolved PSF at
the position of the faint star under consideration. 
We can smooth the image of the star, by correlating it with the 
full PSF $\mathcal{P}$ at the position of the star.
\begin{eqnarray}
Y^{(s)} &=& Y \star \mathcal{P}, \\
Y^{(s)}_{[i,j]} &=& \sum_{k,l}Y_{[i-k,j-l]}\mathcal{P}_{[k,l]},
\end{eqnarray}
where $Y$ is the image of the star, and $Y^{(s)}$ is sometimes called a matched filter. 
A matched filter is a method in which the data $Y$ is correlated (convolved in the 
case of symmetrical PSF) with the PSF $\mathcal{P}$. It is equivalent to optimizing the 
likelihood and therefore provides an optimal map 
where the peak of the map is the likely position of the 
point source (Lang \emph{et al.}, in preparation).

Then, we fit a simple 2d second-order polynomial 
$P(x,y)=a+bx+cy+dx^2+exy+fy^2$ 
to the 3$\times$3 patch centered on the brightest pixel of the
matched-filter image $Y^{s}$.
Upon constructing a universal 9$\times$6 design matrix
\begin{equation}
    \mathbf{A} = 
    \begin{bmatrix}
        1 & x_{1} & y_{1} & x_{1}^{2} & x_{1}y_{1} & y_{1}^{2} \\
        . & . & . & . & . & .  \\
        . & . & . & . & . & .  \\
        . & . & . & . & . & .  \\
        1 & x_{9} & y_{9} & x_{9}^{2} & x_{9}y_{9} & y_{9}^{2}
    \end{bmatrix},
\end{equation}
the free parameters $\{a,b,c,d,e,f\}$
(hereafter compactly denoted by $\mathbf{X}$) can be determined by 
\beq
\mathbf{X} = (\mathbf{A}^{T}\mathbf{A})^{-1}\mathbf{A}^{T}\mathbf{Z},
\label{linearfit}
\eeq
where $\mathbf{Z}$ is given by $(z_{1},...,z_{9})^{T}$,
with $z_{i}$, being the brightness of the $i-$th pixel of the 3$\times$3 patch centered on the brightest pixel of $Y^{(s)}$.
Afterwards, the best fit parameters can be used to compute the centroid coordinate

\beq
  \begin{bmatrix}
      x_{c}\\
      y_{c}\\
  \end{bmatrix} = 
  \begin{bmatrix}
      2d & e\\
      e & 2f\\
  \end{bmatrix}^{-1}
  \begin{bmatrix}
      -b\\
      -c\\
  \end{bmatrix}.
\label{center}
\eeq

It is important to note that the algebraic operation in (\ref{center}) involves 
inverting a 2$\times$2 curvature matrix
\beq
  D = 
  \begin{bmatrix}
      2d & e\\
      e & 2f\\
  \end{bmatrix}.
\eeq

When the curvature matrix $D$ has a zero (or very close to zero) deteminant,
centroid estimates obtained from equation (\ref{center}) can become arbitrarily 
large, which leads to catastrophic outliers. 
In order to tackle this issue, we add a soft regularization term
proportional to $\sigma$ to the diagonals of $D$ prior to inversion.

The procedure of convolving the image of star with the PSF results in a
smoother image. Therefore, a simple second-order polynomial will provide a better fit 
since convolution with the PSF makes the variation of the brightness of the image 
across the 3$\times$3 patch very smooth.

%We should nonlinearly optimize a PSF to saturate the bound.
%But because optimization is done through a chi-squared fitting, 
%this is equivalent to optimizing a matched filter. 
%And if the image is well sampled (in the PSF-convolved image) this
%is equivalent to interpolating a matched filter on a grid. 
%Therefore we expect the matched filter centroiding method to 
%saturate the bound in cases where the image is well sampled.
 
\subsection{Fixed-Gaussian polynomial centroiding}
In the case that we do not know the PSF at the position of star, we change 
the smoothing step in the following way. Instead of smoothing the image 
by convolving it with the PSF, smoothing is done by convolving the image 
with a fixed Gaussian kernel with a fixed size 
\beq
k(\mathbf{x}) = \frac{1}{2\pi w^2}\exp(-\mathbf{x}^{2}/2w^{2}),
\label{eq:gauss}
\eeq
where throughout this study, the full-width at half-maximum of the Gaussian kernel is held at
a fixed value of 2.8 pixels (corresponding to $w \simeq$ 1.2 pixels). The smoothing step is done as follows
\begin{eqnarray}
Y^{(s)} &=& Y \star \mathcal{K}, \\
Y^{(s)}_{[i,j]} &=& \sum_{k,l}Y_{[i-k,j-l]}\mathcal{K}_{[k,l]},
\end{eqnarray}
where $Y$ is the image of the star, $Y^{(s)}$ is the smooth image, and $\mathcal{K}$ is an
array whose elements are given by the Gaussian kernel
\beq
\mathcal{K}_{[k,l]} = k(x_{k},y_{l}).
\eeq
Note that the size of the kernel $\mathcal{K}$ is equal to the size of the kernel $\mathcal{P}$ used in 
the matched-filter polynomial centroiding. Then we apply the same 2d second-order polynomial method (see (\ref{linearfit}), (\ref{center})) to the 3$\times$3 patch centered on the brightest
pixel of the smooth image $Y^{(s)}$. Therefore, for a given star and a smoothing kernel,
the outcome of equation (\ref{linearfit}) can be
plugged into equation (\ref{center}) to find the centroid estimate
of the star. This is inspired by the 3$\times$3 quartic approximation 
used in the \textsl{Sloan Digital Sky Surveys} photometric pipeline \citep{sdss}.

\subsection{Center-of-light centroiding}
In addition to the fitting methods 
mentioned so far, we examine centroiding 
stars by computing their first moments
in a 7$\times$7 patch around the brightest pixel of the image.

\begin{eqnarray}
x_{c} &=& \frac{\sum_{m}x_{m}Y_{m}}{\sum_{m}Y_{m}}, \\
y_{c} &=& \frac{\sum_{m}y_{m}Y_{m}}{\sum_{m}Y_{m}},
\end{eqnarray}
where the summation is done over all the pixels of the 7$\times$7 patch, and $x_{m}$, 
$y_{m}$, and $Y_{m}$, are the $x$ coordinate, $y$ coordinate, and the brightness
of pixel $m$ respectively.

In terms of saturating the Cram\'{e}r-Rao lower bound, we expect this simple 
center-of-light centroiding to perform worse than all other methods mentioned in 
this section. Hereafter, we call this method $7\times7$ moment centroiding.

\section{Tests}\label{sec:data}

We perform two sets of simulations. In the first set, we choose four values of
2, 2.8, 4, and 5.6 pixels for $\gamma$. For each $\gamma$, we generate 100,000 
17 $\times$ 17 postage-stamps of Moffat profiles with centroids randomly drawn
within the central pixel of the 17 $\times$ 17 postage-stamps. Moreover, zero-mean 
uncorrelated Gaussian noise is added to each postage-stamp such that the simulated 
stars are uniformly distributed in log-SNR between SNR = 5 to SNR = 100.

In the second set, we generate 100,000 17$\times$17 postage-stamps
of Moffat profile, with values of $\gamma$ uniformly distributed 
between 2 and 6 pixels, and with centroids drawn randomly within 
the central pixel. We choose four values for SNR: 5, 10, 20, and 40. 
For each SNR, and for each postage-stamp with a given $\gamma$, 
zero-mean uncorrelated Gaussian noise, with standard deviation corresponding 
to SNR and $\gamma$ through equation (\ref{snr2}), is added to each postage-stamp.

In the first experiment, we study how the centroiding error behaves with changing
SNR, while $\gamma$ is held constant. In the second experiment, we study 
how the centroiding error behaves with changing $\gamma$ while SNR is held constant.

\section{Results}\label{sec:result}

\subsection{Experiment 1 : variable SNR; constant $\gamma$}
   
\begin{figure*}[p]~\\
\begin{center}
%\minipage{.8\textwidth}
\includegraphics[width=\linewidth]{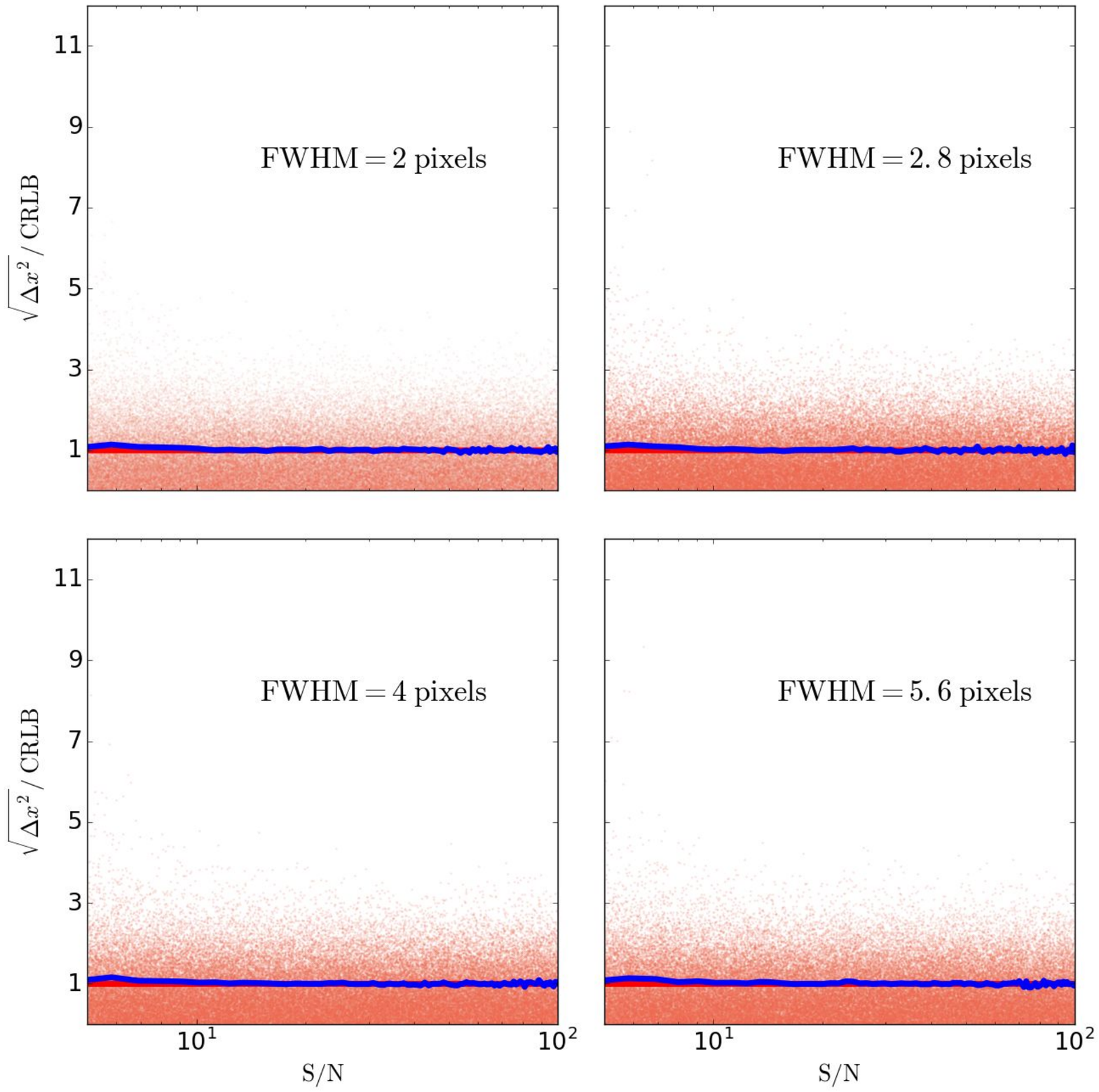}
%\endminipage
\caption{Scatter plots showing the relation between the ratio of error (in x-axis of the centroid poistions) to the CRLB and the signal-to-noise ratio of stars. 
Errors are found from fitting the exact PSF model to the stars,
with FWHM of : 2 (upper left), 2.8 (upper right), 4 (lower left), and 5.6 (lower right)
pixels. In each scatter plot, the blue solid line represents the ratio of the root-mean-squared-error to the CRLB, and the red line represents the ratio achievable by an optimal estimator.}\label{1}
\end{center}
\end{figure*}

\begin{figure*}[p]~\\
\begin{center}
 \includegraphics[width=\linewidth]{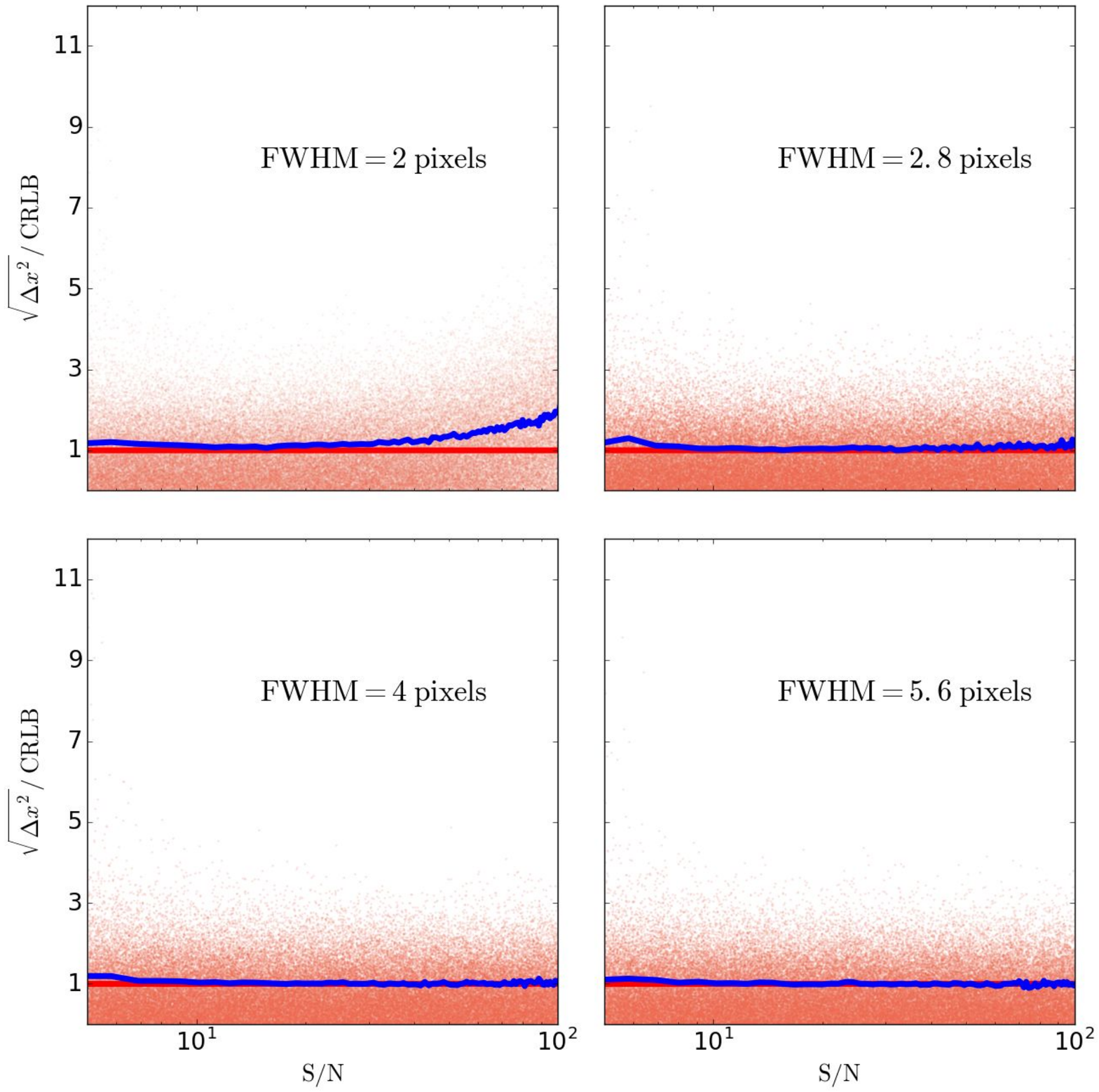}
%\endminipage
 \caption{
 Scatter plots showing the relation between the ratio of error (in x-axis of the centroid poistions) to the CRLB and the signal-to-noise ratio of stars. 
Errors are found from applying the matched filter polynomial centroiding to the stars,
with FWHM of : 2 (upper left), 2.8 (upper right), 4 (lower left), and 5.6 (lower right) pixels. In each scatter plot, the blue solid line represents the ratio of the root-mean-squared-error to the CRLB, and the red line represents the ratio achievable by an optimal estimator.}\label{2}
\end{center}
\end{figure*}

\begin{figure*}[p]~\\
\begin{center}
\includegraphics[width=\linewidth]{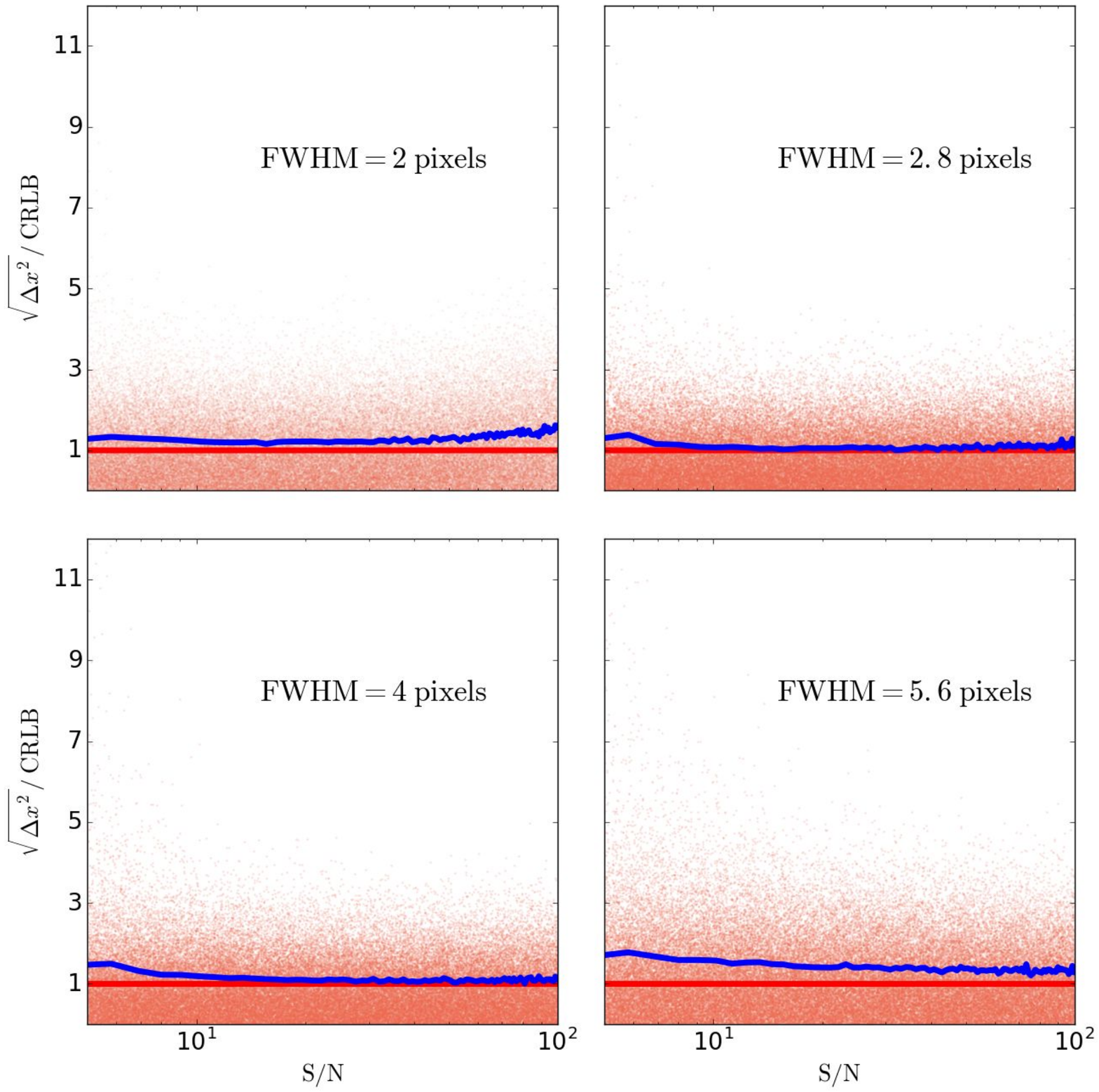}
%\endminipage 
\caption{Scatter plots showing the relation between the ratio of error (in x-axis of the centroid poistions) to the CRLB and the signal-to-noise ratio of stars. 
Errors are found from applying the fixed-Gaussian polynomial centroiding to the stars,
with FWHM of : 2 (upper left), 2.8 (upper right), 4 (lower left), and 5.6 (lower right) pixels. In each scatter plot, the blue solid line represents the ratio of the root-mean-squared-error to the CRLB, and the red line represents the ratio achievable by an optimal estimator.}\label{3}
\end{center}
\end{figure*}

\begin{figure*}[p]~\\
\begin{center}
\includegraphics[width=\linewidth]{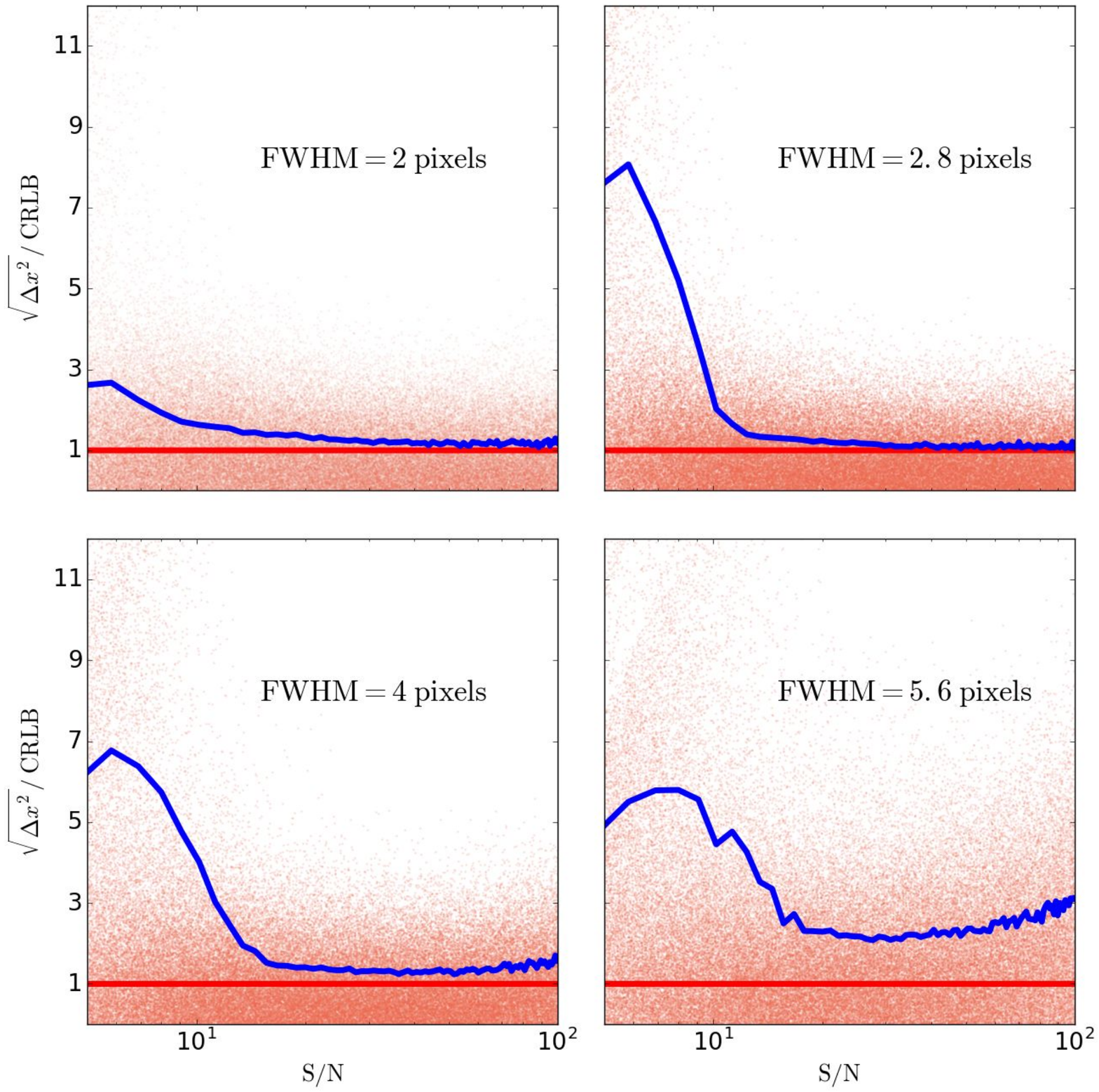}
\caption{Scatter plots showing the relation between the ratio of error (in x-axis of the centroid poistions) to the CRLB and the signal-to-noise ratio of stars. 
Errors are found from applying the 7$\times$7 moment method to the stars,
with FWHM of : 2 (upper left), 2.8 (upper right), 4 (lower left), and 5.6 (lower right) pixels. In each scatter plot, the blue solid line represents the ratio of the root-mean-squared-error to the CRLB, and the red line represents the ratio achievable by an optimal estimator.}\label{4}
\end{center}
\end{figure*}

In this experiment, after finding the centroiding errors for each method,
we compute the RMSE in bins of SNR in order to compare it to the CRLB. 
Results of the first experiment are shown in Figures~\ref{1},~\ref{2},~\ref{3},~\ref{4}. Note that the centroid errors, the CRLB, and the RMSE values shown in these figures are computed for only one component. As we expected, the RMSE from centroiding by fitting the exact PSF model (Figure~\ref{1}) lies on the CRLB. 

%except for simulations with SNR $\la$ 10 where the RMSE gets slightly pulled away
%from the CRLB due to presence of a few outliers. 
Figure~\ref{2} demonstrates that even the matched filter polynomial 
centroiding is able deliver centroiding estimates as efficient as the PSF fitting method in terms of saturating the bound for the simulated stars with $\gamma = 2.8, 4, 5.6$ pixels. For stars with $\gamma$ = 2 pixels, although this method gets very close to saturating the CRLB, the RMSE arising from this method shows slight deviations from the CRLB since the images of stars are not sufficiently smooth even after correlation of these images with the PSF. For simulated images with higher $\gamma$, convolving the data with the PSF results in images that are smooth around the brightest pixel. This enables the polynomial centroiding to deliver estimates that can saturate the CRLB.

The RMSE from the fixed-Gaussian polynomial centroiding (Figure~\ref{3}),
is very close to the CRLB. As we increase $\gamma$ from 2 pixels to 2.8 pixels, RMSE approaches the CRLB. For stars with $\gamma$ = 2 pixels, the rate at which the RMSE from this method drops
eventually becomes smaller than the constant rate at which the CRLB
decreases with increasing SNR. The reason for this is that even after smoothing
the data with a Gaussian kernel, the images are not smooth enough
for a second-order polynomial fitting to deliver estimates with RMSE close to the bound. For stars with $\gamma$ = 2.8 pixels, a significant fraction of information is in the 3$\times$3 patch of the smooth image and this method is able to saturate the bound. When we increase $\gamma$ to 4 and 5.6 pixels, the mismatch between the width of the Gaussian kernel and the PSF increases and the RMSE deviates from the CRLB. The deviation is largest for the simulated stars with $\gamma$ = 5.6 pixels.

On the other hand, Figure~\ref{4} shows that in case of 7$\times$7 moment method, the RMSE becomes quite large as we move toward fainter stars in our simulation.
For stars with larger $\gamma$, centroid estimates from the naive center-of-light 
centroiding do not even come close to saturating the CRLB. As $\gamma$ increases, the 
RMSE deviates further from the CRLB. 

\subsection{Experiment 2 : constant SNR; variable $\gamma$}

\begin{figure*}[p]~\\
\begin{center}
\includegraphics[width=0.65\linewidth]{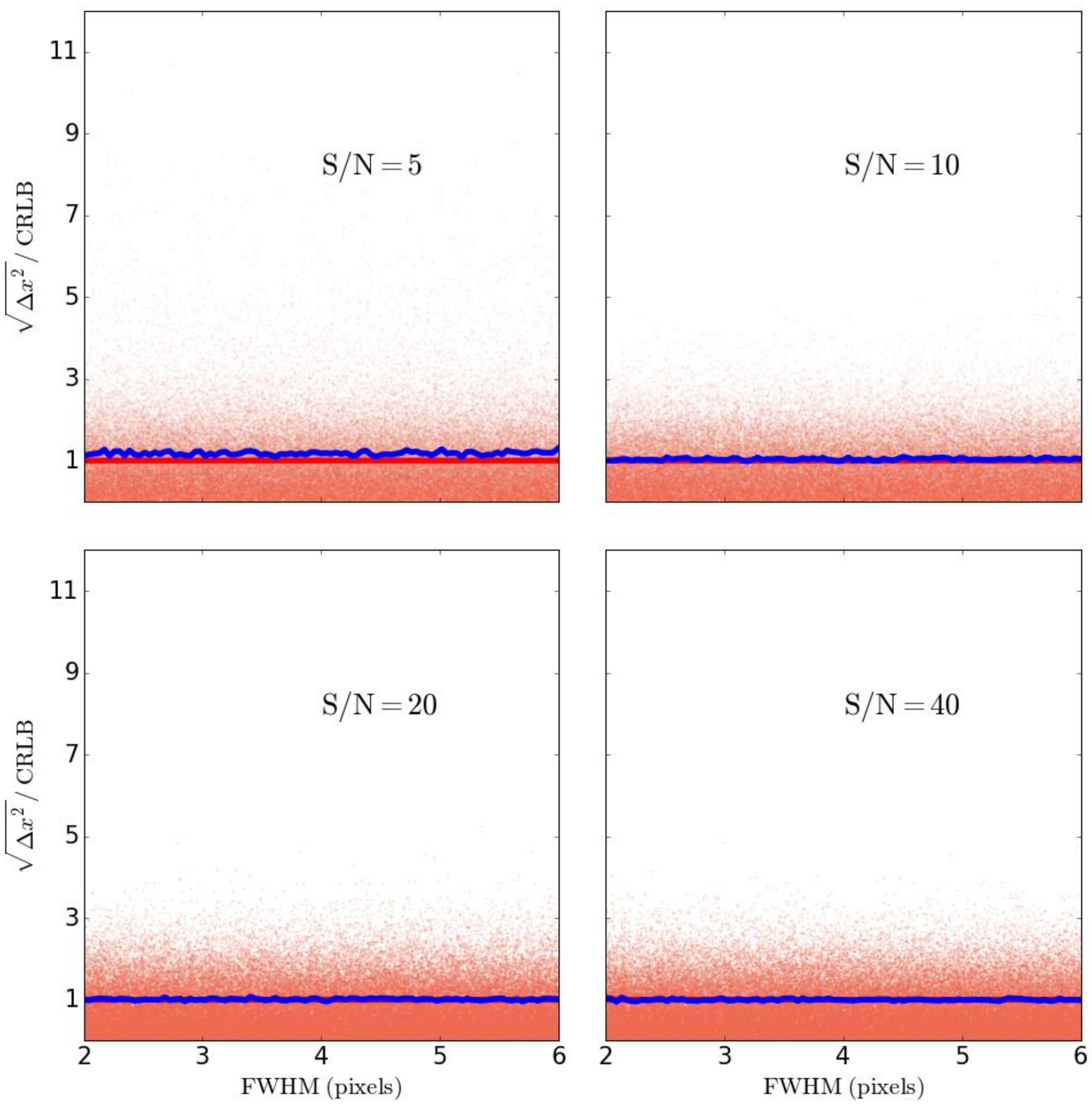}
\caption{Scatter plots showing the relation between the ratio of error (in x-axis of the centroid poistions) to the CRLB and the FWHM of stars.
Errors are found from fitting the exact PSF model to the stars, with SNR  of : 5 (upper left), 10 (upper right), 20 (lower left), and 40 (lower right). In each scatter plot, the blue solid line represents the ratio of the root-mean-squared-error to the CRLB, and the red line represents the ratio achievable by an optimal estimator.}\label{5}
\end{center}
\end{figure*}

\begin{figure*}[p]~\\
\begin{center}
\includegraphics[width=0.65\linewidth]{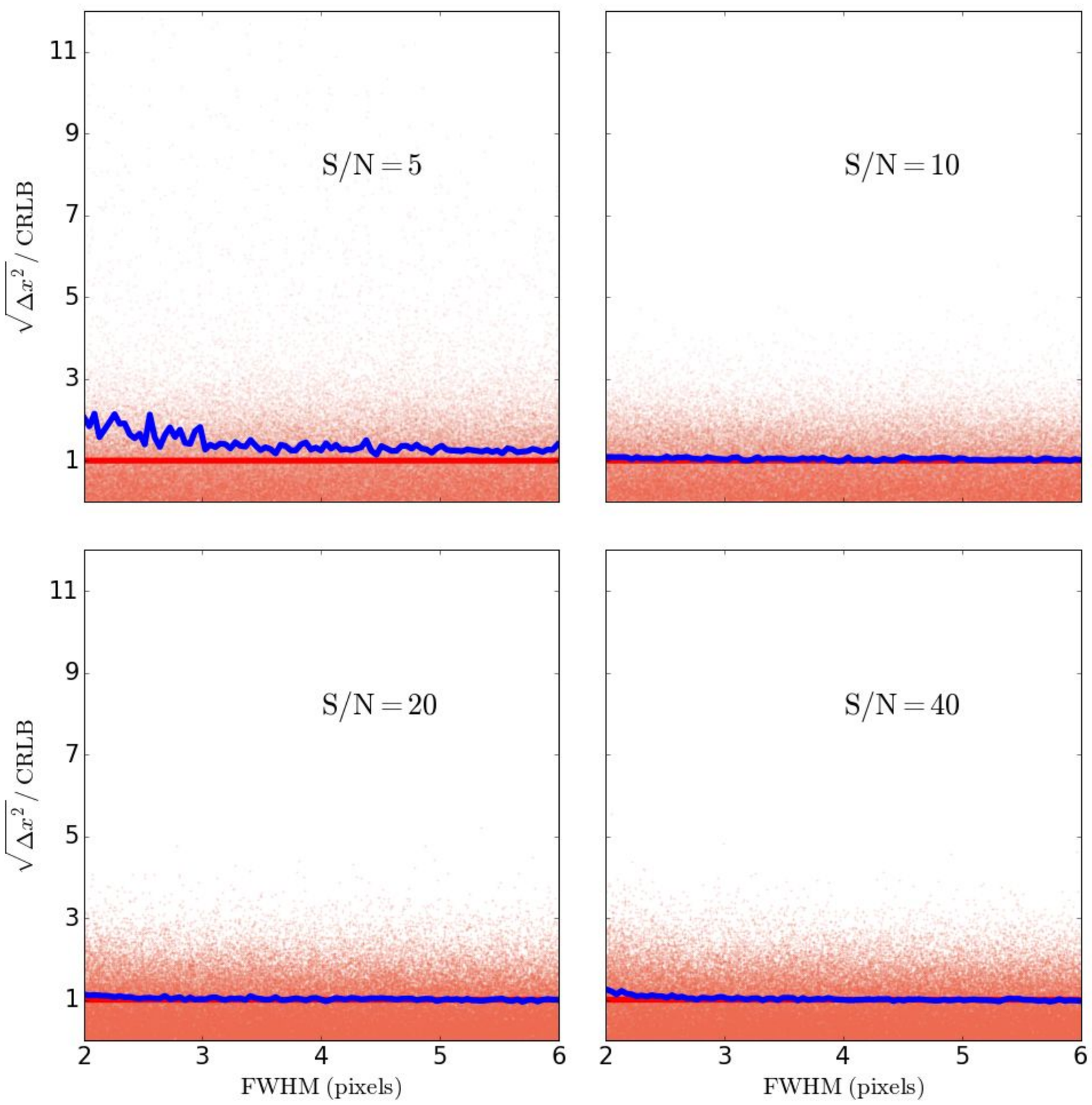}
%\endminipage
\caption{Scatter plots showing the relation between the ratio of error (in x-axis of the centroid poistions) to the CRLB and the FWHM of stars.
Errors are found from applying the matched filter polynomial centroiding to the stars, with SNR  of : 5 (upper left), 10 (upper right), 20 (lower left), and 40 (lower right). In each scatter plot, the blue solid line represents the ratio of the root-mean-squared-error to the CRLB, and the red line represents the ratio achievable by an optimal estimator.}\label{6}
\end{center}
\end{figure*}

\begin{figure*}[p]~\\
\begin{center}
\includegraphics[width=0.65\linewidth]{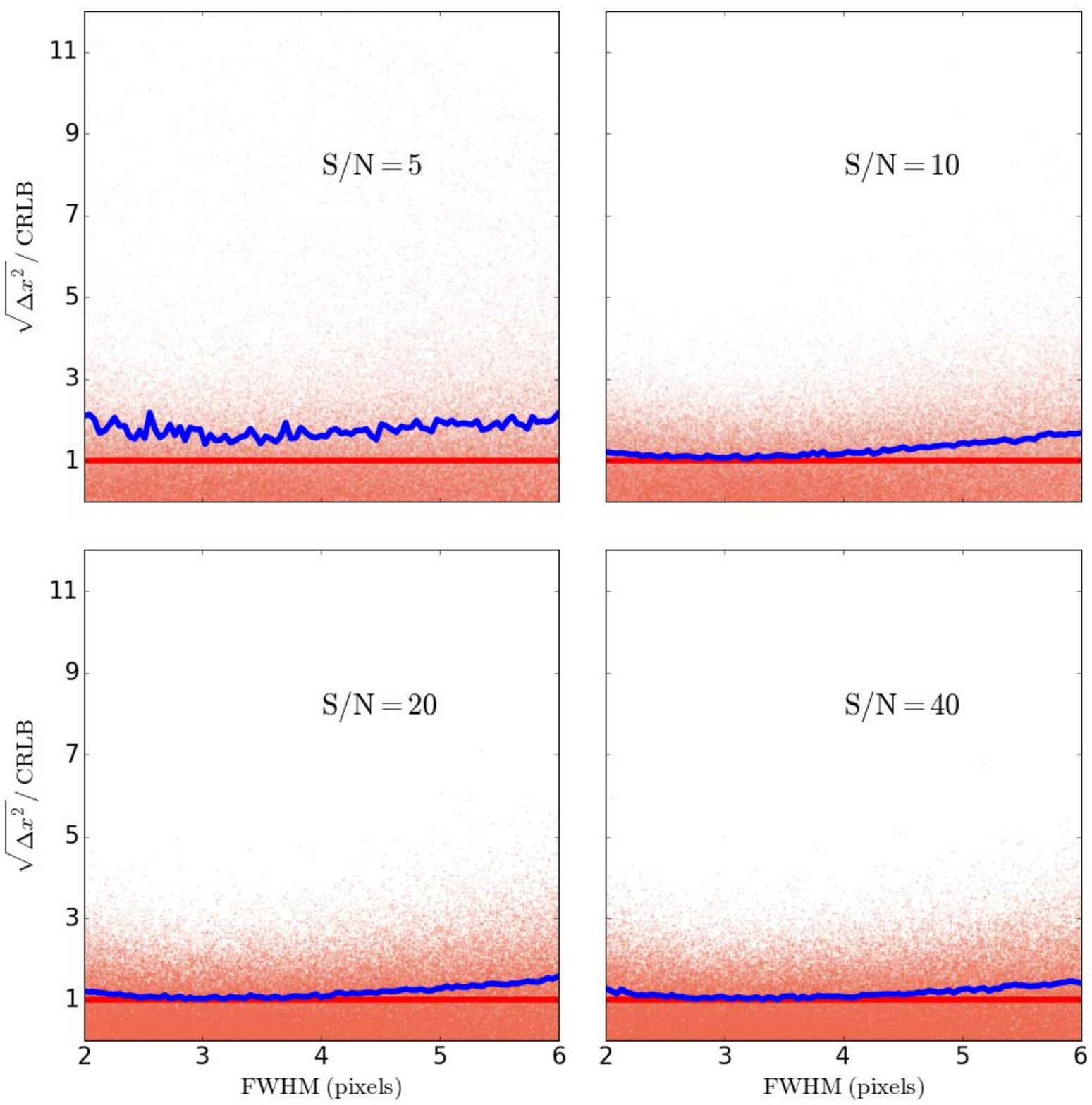}
%\endminipage
\caption{Scatter plots showing the relation between the ratio of error (in x-axis of the centroid poistions) to the CRLB and the FWHM of stars.
Errors are found from applying the fixed-Gaussian polynomial centroiding to the stars, with SNR  of : 5 (upper left), 10 (upper right), 20 (lower left), and 40 (lower right). In each scatter plot, the blue solid line represents the ratio of the root-mean-squared-error to the CRLB, and the red line represents the ratio achievable by an optimal estimator.}\label{7}
\end{center}
\end{figure*}

In this experiment, after finding the centroiding errors for each method, we
compute the RMSE in bins of $\gamma$ in order to compare it to the CRLB. 
Behavior of error as a function of $\gamma$ for different values of SNR,
is shown in Figures~\ref{5},~\ref{6},~\ref{7}, and~\ref{8}. Note that the centroid errors, the CRLB, and the RMSE values shown in these figures are computed for only one component.
 
Once again, the RMSE from centroiding by fitting the exact PSF model as a function of FWHM lies on the CRLB (see Figure~\ref{5}). 
Thus, centroid estimates from fitting the exact PSF model always saturate the CRLB. Once again, we observe that the centroid estimates found by the matched filter polynomial centroiding saturate the CRLB with the
exception of simulated stars with $\gamma$ very close to 2 pixels (see Figure~\ref{6}).

Figure~\ref{7} illustrates that the fixed-Gaussian polynomial method results in RMSE very close to the CRLB. For all four values of SNR, as we increase $\gamma$ from 2 pixels to 3 pixels, the RMSE gets closer to the CRLB since the method starts to perform
slightly better as we move away from undersampled stars and as the FWHM of the smoothing kernel gets closer to that of the simulated images.
After approximately 3 pixels, increasing $\gamma$ results in deviation of the RMSE of the method from the CRLB. This is a characteristic of the fixed-Gaussian polynomial method as we apply it to a smooth image in which some fraction of the available information is lost in the 3$\times$3 patch around the brightest pixel. Furthermore, increasing the SNR from 5 to 40 makes the RMSE (as a function of $\gamma$) closer to the CRLB. 
In the case of extremely faint stars (SNR = 5), the fixed-Gaussian polynomial centroiding fails to saturate the bound.

The centroid estimates obtained from the naive 7$\times$7 moment method (see Figure ~\ref{8}) result in RMSE much larger than the CRLB in all ranges of FWHM and for all four values of SNR in this experiment. 

\section{Discussion}\label{sec:discussion}

An efficient stellar centroiding algorithm must saturate---or come close to saturating---the fundamental 
Cram\'{e}r-Rao lower bound. That is, in all ranges of background noise level, size, radial light profile,
and shape, it must preserve information about the centroids of stars. In practice however,
this is only achievable when we have a reasonably good estimate of the PSF. Since we do not always 
know the exact PSF profile, we must make use of approximate centroiding algorithms. In this work, we
studied how close we get to saturating the CRLB with approximate methods acting on relatively low 
signal-to-noise ratio unsaturated stars.
 
We focused on examples from two classes of centroiding algorithms. The first class contains fast and approximate
methods that do not require any knowledge of the PSF at the positions of stars. Of methods that belong to this class,
we consider centroiding stars based on fitting a second-order
polynomial to a 3$\times$3 patch of star images smoothed by a Gaussian kernel of fixed width, and finding the center of light 
of a 7$\times$7 patch around the brightest pixel of the star.

The second class of centroiding algorithms make use of the PSF (or some good estimate of the PSF)
at the positions of stars. In our investigation, it is assumed that the size and the shape of the PSF are known prior to applying these algorithms to the images of stars. We considered two examples from this class. The first example is the matched filter polynomial centroiding, and the
second example is the PSF fitting. In the PSF fitting method, we find the maximum likelihood estimates of the flux and centroids of stars by fitting a PSF model that has the correct shape and size.  

%The second class of centroiding algorithms contains methods that require knowledge of the PSF (or %having a good estimate
%of the PSF) at the positions of stars. We considered two examples from this class. The first example
%is the matched-filter polynomial centroiding, and the
%second example is the full PSF profile fitting. In terms of saturating the Cram\'{e}r-Rao bound, we %compared the performances of these methods against each other.

In terms of saturating the Cram\'{e}r-Rao bound, we compared the performances of these methods against each other. Our results suggest that in all ranges of FWHM and SNR, the PSF fitting method returns 
centroid estimates that saturate the CRLB. This confirms our expectation that maximum-likelihood estimators saturate the Cram\'{e}r-Rao lower bound. 

\begin{figure*}[p]~\\
\begin{center}
\includegraphics[width=0.65\linewidth]{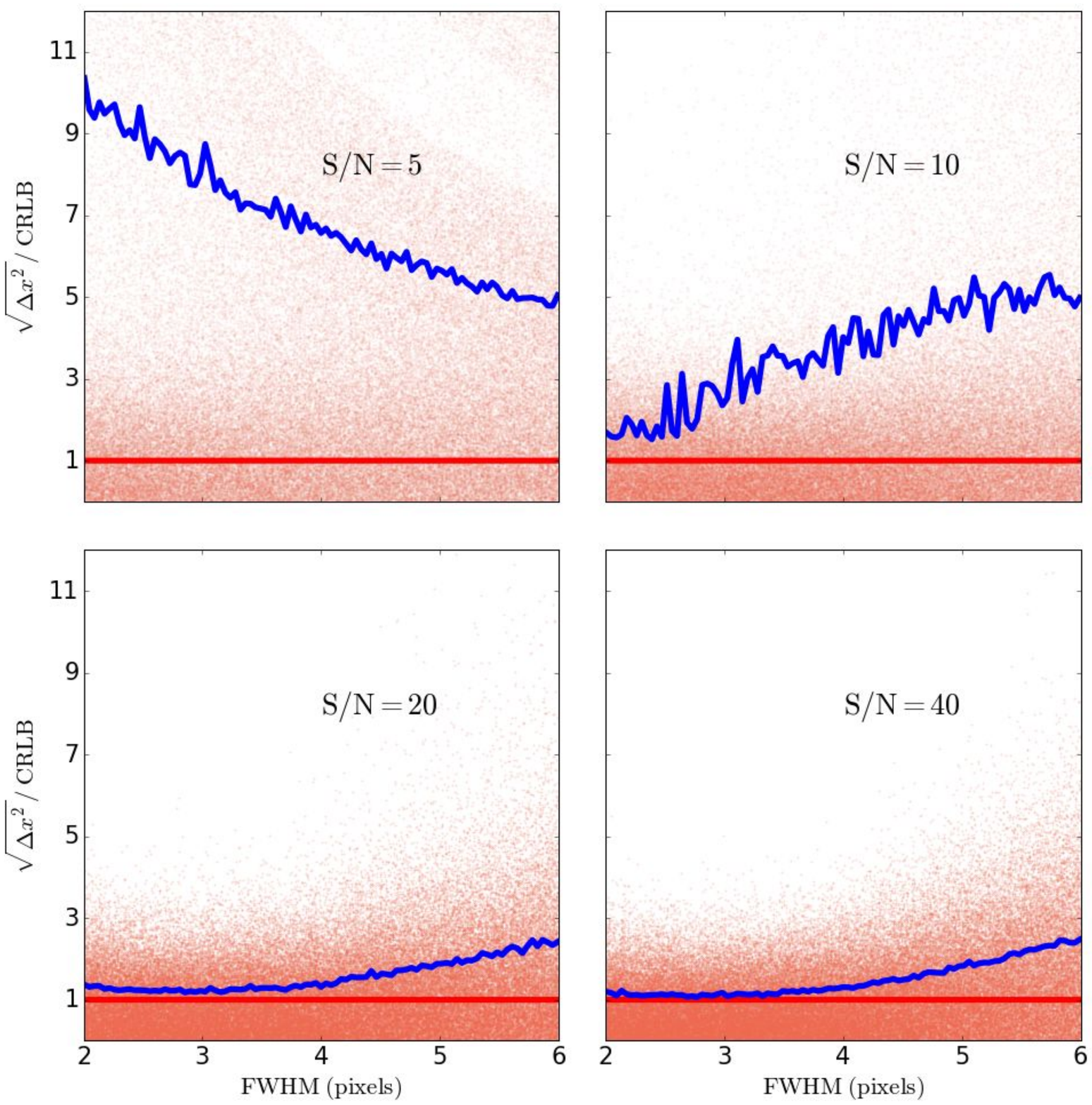}
%\endminipage
\caption{Scatter plots showing the relation between the ratio of error (in x-axis of the centroid poistions) to the CRLB and the FWHM of stars.
Errors are found from applying the 7$\times$7 moment method to the stars, with SNR  of : 5 (upper left), 10 (upper right), 20 (lower left), and 40 (lower right). In each scatter plot, the blue solid line represents the ratio of the root-mean-squared-error to the CRLB, and the red line represents the ratio achievable by an optimal estimator.}\label{8}
\end{center}
\end{figure*}

We note that the estimates found by the 7$\times$7 moment method, except in the case of
very high SNR values and small values of $\gamma$, do not come close to
saturating the CRLB. In a considerable range of PSF sizes and background noise levels, 
this method fails to deliver any centroiding estimate close to saturating the bound. When applied to stars with $\gamma = 2.8, \; 4$ pixels, we find deviation of RMSE from the CRLB as large as $600\%-800\%$ below signal-to-noise ratio of 10. For the simulated stars with $\gamma = 5.6$ pixels, we find deviations as large as $500 \%$ for SNR below 10 and as large as $200 \%$ for SNR $\sim$ 100. It can be noted in Figure~\ref{8} that in the simulations with the lowest SNR (SNR $\sim$ 5), the errors arising from the 7$\times$7 moment method are suppressed by the fact that 17$\times$17 postage-stamps are used to simulate images. Therefore in the case of SNR $\sim$ 5, we expect the deviation of the RMSE from the CRLB to be larger for this method.

On the other hand, the RMSE of centroid estimates of the fixed-Gaussian polynomial centroiding are much closer to saturating the CRLB in all ranges of signal-to-noise ratio even though this method does not require knowledge of the PSF at the positions of stars. We note that when the FWHM of the stars are close to 2.8 pixels (the FWHM of the Gaussian kernel), the fixed-Gaussian polynomial method saturates the CRLB. Deviation of the RMSE of this method from the CRLB is larger for the simulated stars with larger values of FWHM ($\gamma \simeq 5$ pixels). Presence of noise is another limiting factor.
Although this method is able to get very close to saturating the bound in a wide range of
signal-to-noise ratios, it is not reliable in the case of centroiding extremely faint stars ($5<\mathrm{S/N}<10$).
%This is partly due to the fact that in the presence of noise, 
%the brightest pixel of image does not necessarily contain the centroid of stars even %after smoothing 
%the image. 

%The fixed-Gaussian polynomial centroiding technique only takes advantage of the information 
%contained in a 3$\times$3 patch centered on the brightest pixel of the smoothed image which is only 
%well-sampled when the FWHM of the simulated image of star 
%matches that of the smoothing kernel. Thus, when we apply 
%this method to find the centroids of stars with larger FWHM, a certain amount of
%information (encoded in the Cram\'{e}r-Rao lower bound) is lost, and therefore the RMSE
%of these methods deviates from the CRLB. This deviation becomes larger at lower
%signal-to-noise ratios. Besides, the performance of this method slightly degrades
%in the case of undersampled stars (with FWHM close two 2 pixels). 

In matched filter polynomial centroiding, the fixed-Gaussian polynomial method is modified by convolving the image with the correct PSF. Our results on the simulated stars show that the matched filter estimator saturates the CRLB for all PSF sizes and noise levels.  
This is due to the fact that once the images of stars are convolved with the correct PSF, they become smooth that fitting a second-order polynomial to the 3$\times$3 patch centered on the brightest pixel of the smooth image is sufficient for us to obtain results as accurate as those from fitting a PSF profile. 

The Gaussian kernel (see equation \ref{eq:gauss}) in the fixed-Gaussian polynomial centroiding is separable, and correlation of the kernel with an image of star can be performed \emph{exactly} in no time. Therefore in terms of computational cost, this method is more efficient than the matched filter method in which the image of star is correlated with a PSF of arbitrary shape. 

In the case that we have a good estimate of the PSF, the matched filter polynomial method can be faster than PSF fitting method for \emph{centroiding} \emph{purposes}. Additionally, this method is able to saturate the CRLB in a wide range of conditions. It is however important to note that in many cases, reliable estimation of the flux requires a technique as accurate as PSF-fitting. However, in cases in which an investigator only needs an empirical estimates of the centroid offsets, fixed-Gaussian polynomial or matched filter polynomial centroiding methods can be employed with negligible loss of information. For instance in modeling the stellar light curves in the \emph{K2} mission, \citet{dfm} uses a simple polynomial centroiding to marginalize out the systematic trends caused by centroid offsets. 

%However, this method has its own disadvantages. First, we do not %always know the the exact PSF. Second, finding 
%the centroid by profile fitting is computationally expensive, %whereas employing any of the 3$\times$3 polynomial techniques %considered in this study in large scale astronomical surveys %reduces the computational cost of initial astrometry of the %point sources considerably. 

Moreover, we note that the PSF fitting method can be made faster by only keeping the term proportional to the dot product of the PSF model and the image in $\chi^2$:
\beq
\chi^{2} = \big(\mathbf{y} \cdot \mathbf{y} -2F\mathbf{y} \cdot \mathbf{m} + F^{2}\mathbf{m} \cdot \mathbf{m}\big)/\sigma^{2},
\eeq
where $F$ is the flux, $\sigma$ is the per-pixel uncertainty, the dot product between two vectors is denoted by $(\cdot)$, and the image of star and the normalized shifted PSF model are denoted by $\mathbf{y}$ and $\mathbf{m}$ respectively. Upon varying only the centroid, the terms $\mathbf{y}.\mathbf{y}$ and $\mathbf{m}.\mathbf{m}$ remain approximately constant. However, this only allows us to vary the position of centroid, and not the flux, while fitting the PSF model to the star. 

Finding a centroid coordinate that maximizes the dot product of the PSF and the star image is equivalent to finding the peak of the correlation of the PSF and the image. Therefore optimizing the modified $\chi^2$ is equivalent to finding the location of the peak of the matched filter. 

In the initial smoothing step of the fixed-Gaussian polynomial method, the image of the star is
correlated with an approximate Gaussian PSF. When there is mismatch between the widths of the smoothing kernel and the that of the PSF, we loose some information by employing a 3$\times$3 polynomial fitting. When we have the advantage of knowing the PSF, this issue can be resolved by employing the matched filter polynomial method.

In this investigation we showed that PSF fitting always performs better---in terms of saturating the CRLB---at centroiding stars. Having a reasonable PSF model always helps us obtain more reliable centroid estimates, but over a certain range of low signal-to-noise ratios and PSF sizes, one can achieve sensibly accurate results by employing a simple 3$\times$3 method after smoothing the image with a Gaussian kernel of a fixed width, and without making any assumption about the PSF model 
at the positions of stars.

In this investigation we narrowed our focus on a set of data simulated from a particular PSF profile. Although there are various cases where Moffat profiles provide reasonable representations of the point spread function, these profiles are not generic enough to let us reach a more general conclusion.

This work was partially supported by the NSF (grants IIS-1124794 and AST-1517237), NASA (grant NNX12AI50G), and the Moore-Sloan Data Science Environment at NYU. We thank Jo Bovy and Alex Malz for discussions related to this work. We are also grateful to Dustin Lang and Alex Malz for reading and making comments on draft.

%\begin{thebibliography}{70}
%\bibitem[Bertin \& Arnouts (1996)]{sextractor} Bertin, E., 5Arnouts, S., 1996,  A\&AS , 117, 393
%\bibitem[Cram\'{e}r (1946)]{cramer} Cram\'{e}r, H., 1946, %Mathematical methods of statistics, Princeton university press
%\bibitem[Le Cam (1953)]{lecam} Le Cam, L. M., 1953, University of 5California publications in statistics, Vol. 1 (no. 11.), 277
%\bibitem[Lobos \etal (2015)]{lobos} Lobos, R.~A., Silva, J.~F., %Mendez, R.~A., Orchard, M., 2015, \pasp , 127, 1166
%\bibitem[Lupton \etal (2001)]{sdss} Lupton, R. \etal, 2001,  %arXiv:astro-ph/0101420
%\bibitem[Schechter \etal (1993)]{dophot} Schechter, P. L., Mateo, %M., Saha, A., 1993, \pasp , 105, 1342
%\bibitem[Stetson (1987)]{daophot} Stetson, P. B., 1987, \pasp , %99, 191
%\bibitem[Trujillo etal. (2001)]{moffat} Trujillo, I., \etal, %2001, \mnras , 328, 977
%\end{thebibliography}

\clearpage

%%%%%%%%%%%%%%%%%%%%%%%%%%%%%%%%%%%%%%%%%%%%%%%%%%%%%%%%%%%%%%%%%%%%%%%%%%% FWHM PLOTS %%%%%%%%%%%%%%%%%%%%%%%%%%%%%%%%%%%%%%%%%%

\bibliographystyle{yahapj}
%\begin{bibliography}
\bibliography{centroid}

\begin{thebibliography}{}
\expandafter\ifx\csname natexlab\endcsname\relax\def\natexlab#1{#1}\fi

\bibitem[{{Doob}(1946)}]{crlb}
{Doob}, J.~L. 1946, Science, 104, 450

\bibitem[{{Foreman-Mackey} {et~al.}(2015){Foreman-Mackey}, {Montet}, {Hogg},
  {Morton}, {Wang}, \& {Sch{\"o}lkopf}}]{dfm}
{Foreman-Mackey}, D., {Montet}, B.~T., {Hogg}, D.~W., {et~al.} 2015, \apj, 806,
  215

\bibitem[{Le~Cam(1953)}]{lecam}
Le~Cam, L.~M. 1953 (University of California press)

\bibitem[{{Lobos} {et~al.}(2015){Lobos}, {Silva}, {Mendez}, \&
  {Orchard}}]{lobos}
{Lobos}, R.~A., {Silva}, J.~F., {Mendez}, R.~A., \& {Orchard}, M. 2015, \pasp,
  127, 1166

\bibitem[{{Lupton} {et~al.}(2001){Lupton}, {Gunn}, {Ivezi{\'c}}, {Knapp}, \&
  {Kent}}]{sdss}
{Lupton}, R., {Gunn}, J.~E., {Ivezi{\'c}}, Z., {Knapp}, G.~R., \& {Kent}, S.
  2001, in Astronomical Society of the Pacific Conference Series, Vol. 238,
  Astronomical Data Analysis Software and Systems X, ed. F.~R. {Harnden}, Jr.,
  F.~A. {Primini}, \& H.~E. {Payne}, 269

\bibitem[{{Rowe} {et~al.}(2015){Rowe}, {Jarvis}, {Mandelbaum}, {Bernstein},
  {Bosch}, {Simet}, {Meyers}, {Kacprzak}, {Nakajima}, {Zuntz}, {Miyatake},
  {Dietrich}, {Armstrong}, {Melchior}, \& {Gill}}]{galsim}
{Rowe}, B.~T.~P., {Jarvis}, M., {Mandelbaum}, R., {et~al.} 2015, Astronomy and
  Computing, 10, 121

\bibitem[{{Trujillo} {et~al.}(2001){Trujillo}, {Aguerri}, {Cepa}, \&
  {Guti{\'e}rrez}}]{moffat}
{Trujillo}, I., {Aguerri}, J.~A.~L., {Cepa}, J., \& {Guti{\'e}rrez}, C.~M.
  2001, \mnras, 328, 977

\bibitem[{{Zuntz} {et~al.}(2013){Zuntz}, {Kacprzak}, {Voigt}, {Hirsch}, {Rowe},
  \& {Bridle}}]{im3shape}
{Zuntz}, J., {Kacprzak}, T., {Voigt}, L., {et~al.} 2013, \mnras, 434, 1604

\bibitem[{{Zuntz} {et~al.}(2014){Zuntz}, {Kacprzak}, {Voigt}, {Hirsch}, {Rowe},
  \& {Bridle}}]{im3shape_code}
---. 2014, {IM3SHAPE: Maximum likelihood galaxy shear measurement code for
  cosmic gravitational lensing}, Astrophysics Source Code Library,
  ascl:1409.013

\end{thebibliography}
%\end{bibliography}
\end{document}